# Borophene as an extremely high capacity electrode material for Li-ion and Na-ion batteries

*Xiaoming Zhang*,[a,b] *Junping Hu*,[c] *Yingchun Cheng*,[d] *Hui Ying Yang*,[*a] *Yugui Yao*,[*b] *Shengyuan A. Yang*[*a]

[a]Research Laboratory for Quantum Materials and Engineering Product Development Pillar, Singapore University of Technology and Design, Singapore 487372, Singapore. E-mail: yanghuiying@sutd.edu.sg; shengyuan_yang@sutd.edu.sg
[b]School of Physics, Beijing Institute of Technology, Beijing 100081, China. E-mail: ygyao@bit.edu.cn
[c]School of Science, Nanchang Institute of Technology, Nanchang 330099, China
[d]Key Laboratory of Flexible Electronics & Institute of Advanced Materials, Jiangsu National Synergetic Innovation Center for Advanced Materials, Nanjing Tech University, Nanjing 211816, China

**ABSTRACT:** "Two-dimensional (2D) materials as electrodes" is believed to be the trend for future Li-ion and Na-ion batteries technologies. Here, by using first-principles methods, we predict that the recently reported borophene (2D born sheet) can serve as an ideal electrode material with high electrochemical performance for both Li-ion and Na-ion batteries. The calculations are performed on the two experimentally stable borophene structures, namely $\beta_{12}$ and $\chi_3$ structures. The optimized Li and Na adsorption sites are identified, and the host materials are found to maintain good electric conductivity before and after adsorption. Besides advantages including small diffusion barriers and low average open-circuit voltages, most remarkably, the storage capacity can be as high as 1984 mA h $g^{-1}$ in $\beta_{12}$ borophene and 1240 mA h $g^{-1}$ in $\chi_3$ borophene for both Li and Na, which is several times higher than the commercial graphite electrode and is the highest among all the 2D materials discovered to date. Our results highly support that borophenes can be appealing anode materials for both Li-ion and Na-ion batteries with extremely high power density.

# 1 Introduction

Energy storage has been the technology that powers the world, with applications ranging from portable electronic devices to electric vehicles, and to large-scale power grid systems. Among different types of energy storage systems, secondary batteries have received remarkable attention because of their great advantage of compact size and high efficiency[1-2]. Especially, rechargeable lithium-ion batteries (LIBs), have been widely used since their first commercialization in 1991, due to their excellent combination of outstanding reversible capacity, high power density, and long cycle life[3-6]. Even though, to meet the increasing demand for the rapid development of the electronic market nowadays, the efforts of seeking for new LIBs with enhanced capacity have never diminished[7-9]. One current research focus is to explore advanced electrode materials, because the capability of LIBs greatly depends on the performances of their electrode materials. Compared with the rapid development of cathode materials, the innovation of anode materials is far from satisfactory, which have been mostly limited to carbon-related materials. As the most widely used LIBs anode, graphite possess good cycling stability and low cost[10], but is hampered from further developments by its low capacity and poor rate capability[11]. Therefore, exploiting suitable anode materials with good performance, especially with larger capacity, is urgently needed. Besides LIBs, Na-ion batteries (NIBs) have also received great research interest in recent years[12-14], mostly driven by their superiority of low cost and high operating safety. Similar to LIBs, developing high capacity electrodes for NIBs is also of great significance.

Two-dimensional (2D) materials are believed to be of good choice as electrodes for metal ion batteries, because their unique morphology enables fast ion diffusion and provides maximum ion insertion channels with fully surface exposed[15]. Up to now, many 2D materials are predicted to show good performance as electrodes for both LIBs and NIBs, including: (i) graphene and related systems[16-18]; (ii) transition-metal dichalcogenides, such as $MoS_2$[19], $WS_2$[20], and $VS_2$[21]; (iii) transition-metal carbides (also known as MXenes), such as $Mo_2C$[22], $Ti_3C_2$[23], and $V_2C$[24]; (iv) other 2D systems, such as metal nitrides[25], metal oxides [26] and so on[27-29]. Among these 2D electrodes materials, the maximum capacity for LIBs or NIBs mostly locates at the range of 300-600 mA h g$^{-1}$, which already show obvious superiority compared with the commercial graphite electrodes (372 mA h g$^{-1}$ [11]). Nevertheless, continued efforts have been made to explore novel 2D materials, chasing for record-high storage capacities.

Very recently, a new type of 2D material, borophene (2D born sheet) has been successfully grown on the Ag(111) substrates[30,31], with inherently metallic conductivity. Upon different deposition temperatures, experiments yield two types of borophene: one is a striped phase, the other is a homogeneous phase. For the striped borophene, a former experiment[30] attribute it to a buckled triangular lattice without vacancy (here, denoted as $_\triangle$Borophene); however, the latest experimental[31] and computational[32-34] evidences indicate that the $\beta_{12}$ borophene sheet [see Fig. 1(a)] with 1/6 vacancies is better energetically favored during the nucleation and growth of the striped borophene. One the other hand, the structure of the homogeneous borophene is experimentally determined which takes the $\chi_3$ structure with a 1/5 vacancy concentration[30,31], as shown in Fig. 1(c). Up to now, borophene has been found to possess many extraordinary properties, such as strongly anisotropic mechanical properties[35], phonon-mediated superconductivity[36,37], and novel magnetism and metal-semiconductor-metal transition[38]. Two recent works[39,40], which systematically investigated the electronic, optical, thermodynamic and mechanical properties of borophene are also notable. Particularly, considering its excellent metallic character and low mass density, one may naturally wonders: Can borophene be a superior electrode material for Li-ion or Na-ion batteries? Before the experimental realization of borophene, the calculations by Banerjee and co-workers have already proposed the great possibility of $\alpha_1$ 2D boron sheet for the application of LIB[41], even though the atomic configuation of $\alpha_1$ phase is quite different from the experimental $\beta_{12}$ and $\chi_3$ borophene. Very recently, Jiang *et al.* report $_\triangle$Borophene possess high capacity for Li-storage[42];

however, as discussed above, the $_\triangle$Borophene is in fact not reasonable to depict the striped phase of borophene observed in experiments. So far how the two experimentally stable borophenes (i.e., $\beta_{12}$ (the striped phase) and $\chi_3$ (the homogeneous phase) borophenes) behave as battery electrodes is still unknown.

Motivated by the previous discussions, in the current work, based on first-principles calculations, we investigate the properties of borophene that are central to its performance as battery electrode materials. Our study is focused on the two experimentally realized borophene structures, i.e., the striped $\beta_{12}$ and the homogeneous $\chi_3$ structures. We firstly perform full geometry optimization for the structures of the host $\beta_{12}$ and $\chi_3$ borophene, then study the Li/Na adsorption and diffusion processes. We show that both Li and Na can be adsorbed on borophenes with good metallic conductivities maintained, meanwhile both low surface diffusion barrier and low average open-circuit voltage are obtained. Most remarkably, we find that the theoretical capacity for Li/Na can reach 1984 mA h g$^{-1}$ for $\beta_{12}$ borophene and 1240 mA h g$^{-1}$ for $\chi_3$ borophene, which is 3 to 5 times higher than the commercial graphite electrodes and is the highest among all the 2D materials studied to date. Our results suggest borophenes are highly promising to serve as anode for both LIBs and NIBs with extremely high capacity.

## 2 Computational methods

Our first-principles calculations are based on the density functional theory (DFT), and are performed in conjunction with the projector-augmented wave (PAW) method[43], as implemented in the Vienna Ab initio Simulation Package (VASP)[44]. The generalized gradient approximation (GGA)[45] with the Perdew-Burke-Ernzerhof (PBE)[46] functional is used for the exchange-correlation. The cutoff energy is chosen as 500 eV for the plane-wave expansion. To avoid artificial interactions between the film and the periodic images, a vacuum space with a thickness of 20 Å is used. Both the atomic positions and lattice constants are fully relaxed. The convergence criteria for energy and force are set to be $10^{-5}$ eV and 0.01 eV Å$^{-1}$, respectively. The long-range van der Waals interactions are taken into account by using the DFT-D2 method. For $\beta_{12}$ borophene, Monkhorst-Pack k-point meshes[47] with sizes of $18 \times 12 \times 1$ and $30 \times 20 \times 1$ were used for geometrical optimization and static electronic structure calculation, respectively; for $\chi_3$ borophene, the corresponding k-point meshes are chosen as $15 \times 15 \times 1$ and $21 \times 21 \times 1$.

# 3  Results and discussions

**Crystal and electronic structures of borophene**

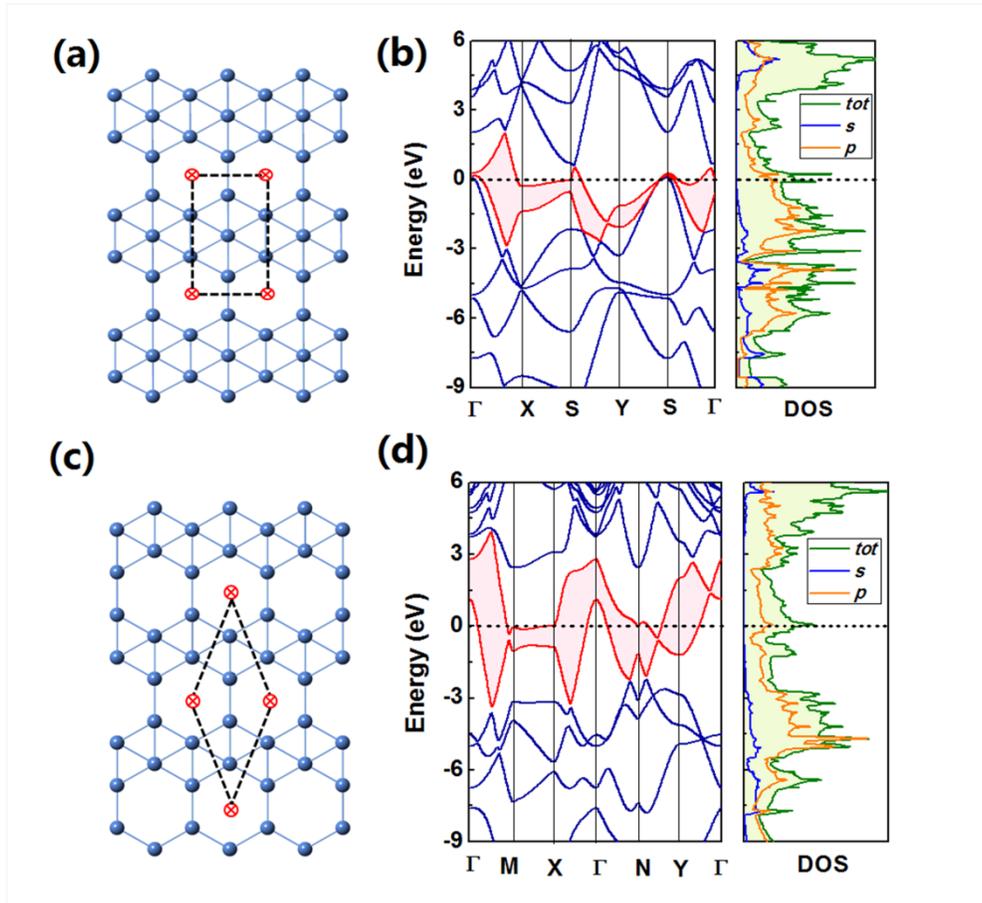

**Fig. 1** Atomic and electronic structures of borophenes. (a) and (b) are given for $\beta_{12}$ borophene, and (c) and (d) for $\chi_3$ borophene. In (a) and (c), the cyan balls and red circles represent the boron atoms and vacancy sites, and the black dashed lines profile the unit cells for $\beta_{12}$ and $\chi_3$ borophene. In (b) and (d), the bands with crossing the Fermi level are highlighted as red shadows.

To begin with, we first show the relaxed geometrical structures of both types of borophenes in Fig. 1(a) and 1(c). As depicted in the figures, vacancy sites exhibits a rectangular pattern for $\beta_{12}$ borophene, and a rhombic pattern for $\chi_3$ borophene. Besides, the vacancy concentration also differs between the two structures: the ratio between vacancy sites and boron sites is 1:5 for $\beta_{12}$ borophene and 1:4 for $\chi_3$ borophene. For $\beta_{12}$ borophene, the equilibrium lattice parameters are: a = 2.926 Å and b =5.068 Å; for $\chi_3$ borophene, the optimized lattice parameter is a=b=4.490 Å, with a rhombic cell angle as 38.181°. These values are consistent with previous results[31,33]. The electronic band structures of $\beta_{12}$ and $\chi_3$ borophenes with optimized lattice parameters are shown in Fig. 1(b) and 1(d). For both structures, as highlighted in the figures, there are two bands crossing the Fermi level, exhibiting a metallic band character. From the calculated density of states (DOS), we find that the *s* orbitals have little contributions to the electronic states near the Fermi level, and the metallic characters of borophenes are mostly dominated by the *p* orbitals. The unambiguous metallicity of borophene makes it possible to be used as a battery electrode material.

**Li/Na adsorption and diffusion on borophene**

To investigate the Li/Na adsorption on borophenes, we first examine the most favorable adsorption sites for an isolated Li or Na atom. Herein, the calculations are performed using a 2 × 2 × 1 supercell for both borophene structures. When one Li or Na atom is adsorbed, the corresponding chemical stoichiometry is $B_{20}Li_1/B_{20}Na_1$ for $\beta_{12}$ borophene, and $B_{16}Li_1/B_{16}Na_1$ for $\chi_3$ borophene. Considering the symmetry of borophenes lattices, there exist eleven typical adsorption sites (denoted as $S_\beta 1 - S_\beta 11$) for $\beta_{12}$ borophene and five (denoted as $S_\chi 1 - S_\chi 5$) for $\chi_3$ borophene, which are schematically shown in Fig. 2(a) and 2(b). After structural optimizations of the above adsorption configurations, we indentify the most favorable adsorption sites of Li/Na ion by computing the adsorption

energies, defined as

$$E_{ad}(\beta_{12}) = E_{B_{20}M_1} - E_{B_{20}} - E_M \quad (1)$$

$$E_{ad}(\chi_3) = E_{B_{16}M_1} - E_{B_{16}} - E_M \quad (2)$$

where $M=$ (Li, Na), $E_M$ is the cohesive energy of the bulk metal (Li/Na); $E_{B_{20}}$ (or $E_{B_{16}}$) and $E_{B_{20}M_1}$ (or $E_{B_{16}M_1}$) are the total energies of $\beta_{12}$ (or $\chi_3$) borophene before and after metal-ion adsorption. A negative value of adsorption energy means that the corresponding atom prefers to adsorb on the host materials instead of forming a metal cluster.

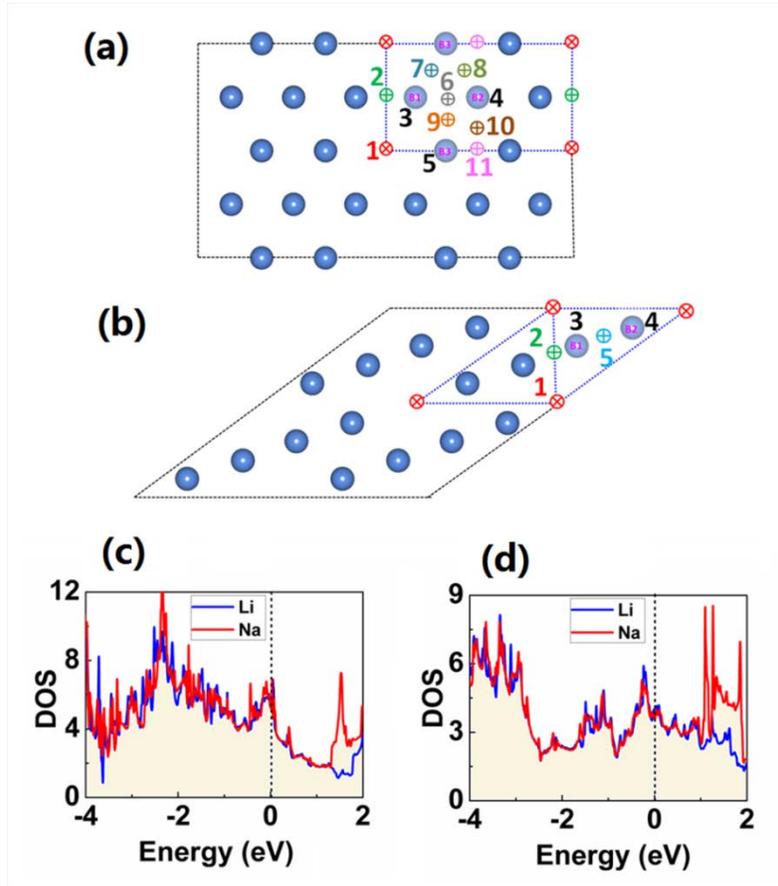

**Fig. 2** (a) The considered Li/Na adsorption sites on the surface of $\beta_{12}$ borophene (top view). $S_\beta 1$, $S_\beta 3$, $S_\beta 4$, and $S_\beta 5$ are on the top of vacancy, B-1, B-2 and B-3, respectively. $S_\beta 2$, $S_\beta 6$, $S_\beta 7$, $S_\beta 8$, and $S_\beta 11$ are on the top of the bridge sites. $S_\beta 9$ and $S_\beta 10$ are on the top of the center of the triangle. (c) Density of states (DOS) for Li- and Na-adsorbed $\beta_{12}$ borophene. (b) and (d) are similar to (a) and (c), but for the case of $\chi_3$ borophene.

For $\beta_{12}$ borophene, during the geometrical optimization, we find that the adsorbed Li/Na at $S_\beta 5$-$S_\beta 8$ sites would automatically shift to $S_\beta 1$ site, because the energy barrier between the configurations of Li/Na adsorbing on $S_\beta 5$-$S_\beta 8$ and $S_\beta 1$ sites are extremely small. Similarly, the adsorbed Li/Na at $S_\beta 3$ would shift to $S_\beta 2$ site, and $S_\beta 9$-$S_\beta 10$ shift to $S_\beta 11$ site. So there are only four (meta-)stable adsorption sites: $S_\beta 1$, $S_\beta 2$, $S_\beta 4$, and $S_\beta 11$. The calculated Li and Na adsorption energies for these sites are listed in Table I. From the adsorption energies, we find that: (i) all the adsorption energies are negative irrespective of adsorption sites for both Li and Na atoms on $\beta_{12}$ borophene, indicating the Li/Na storage is exothermic and spontaneous, which is the prerequisite for LIB and NIB applications; (ii) the most energetically favorable adsorption sites for both Li and Na are located above the vacancies (i.e. the $S_\beta 1$ site), while other three adsorption sites ($S_\beta 2$, $S_\beta 4$ and $S_\beta 11$) have similar adsorption energies that are much higher than that of $S_\beta 1$. For the case of $\chi_3$ borophene, upon geometrical optimization, the adsorbed Li/Na at $S_\chi 3$-$S_\chi 5$ sites will shift to $S_\chi 1$ site. As the results, there are two (meta-)stable adsorption sites: $S_\chi 1$ and $S_\chi 2$. The obtained adsorption energies are also listed in Table I. Therefore, in $\chi_3$ borophene, the most stable Li/Na adsorption is also situated above the vacancy sites (i.e. $S_\chi 1$ site), which is similar to the case of $\beta_{12}$ borophene.

**Table 1** Adsorption energies for Li/Na on different sites of $\beta_{12}$ and $\chi_3$ borophenes.

|  | $S_\beta 1$ (eV) | $S_\beta 2$ (eV) | $S_\beta 4$ (eV) | $S_\beta 11$ (eV) |  | $S_\chi 1$ (eV) | $S_\chi 2$ (eV) |
|---|---|---|---|---|---|---|---|
| $\beta_{12}$-B + Li | -1.766 | -1.088 | -1.025 | -1.046 | $\chi_3$-B + Li | -1.433 | -0.803 |
| $\beta_{12}$-B + Na | -1.487 | -1.163 | -0.987 | -1.027 | $\chi_3$-B + Na | -1.230 | -0.878 |

Further, we carry out the Bader charge analysis and electronic structure calculations for the most stable Li/Na adsorption configurations for $\beta_{12}$ and $\chi_3$ borophene. During the adsorption process, the Li atom is found to transfer ~0.86 e to both $\beta_{12}$ and $\chi_3$ borophene, and the Na atom transfers ~ 0.82 e. This demonstrates that both Li and Na atoms can be chemically adsorbed on borophenes and they form chemical compounds, corresponding to the redox reaction in electrode materials during the battery operation. Then we turn to the DOS for $\beta_{12}$ and $\chi_3$ borophenes after Li/Na adsorption, as shown in Fig. 2(c) and 2(d). We find that the DOS profile near the Fermi level is not affected much upon Li or Na adsorption for both structures. More importantly, both $\beta_{12}$ and $\chi_3$ borophenes maintain their metallic nature after the adsorption of Li or Na atom with sizable DOS locating at Fermi level. The metallic character for both pristine borophenes and their Li/Na intercalated states ensures good electronic conductivity, which paves the way for their applications as battery electrodes.

It is well-known that, the diffusion barrier, which determines charge−discharge rate of rechargeable batteries, is another key parameter to evaluate the performance of electrodes. To study the diffusion properties of Li/Na on borophene, we examine the optimal diffusion paths and calculate the corresponding diffusion barriers by using the climbing-image nudged elastic band (NEB) method[48]. For $\beta_{12}$ borophene, as initial configurations, we consider the three pathways with high structural symmetry between the neighboring most stable adsorption sites. The top view of these migration pathways (denoted as P1, P2, and P3) are schematically shown in Fig. 3(a). After calculation, the obtained energy barriers for Li and Na on $\beta_{12}$ borophene are shown in Fig. 3(b), where the saddle points are found for each path. Obviously, for both Li and Na, the path of P3 show the highest diffusion barrier, while those of P1 and P2 are much lower, which is consistent with the previous estimations of the adsorption energy. It should also be noticed that, P1 and P2 showing quite similar energy in the middle of the diffusion path, which respectively correspond to the adsorption sites of $S_\beta 2$ and $S_\beta 3$, consistent with the fact that the Li/Na atom at $S_\beta 3$ site will automatically shift to $S_\beta 2$ site during geometrical optimizations. As a result, the calculated lowest diffusion barrier on the surface of $\beta_{12}$ borophene is 0.66 eV for Li ion and 0.33 eV for Na ion. For $\chi_3$ borophene, the choice of the initial migration pathways (denoted as P1, P2, and P3) and the corresponding energy barriers are shown in Fig. 4(a) and 4(b), respectively. It is found that, P1 has the lowest diffusion barrier on the surface of $\chi_3$ borophene with a value of 0.60 eV for Li ion and 0.34 eV for Na ion.

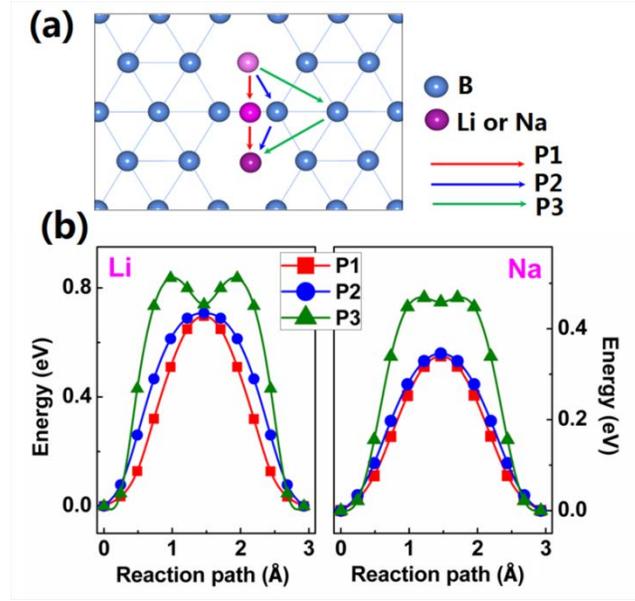

**Fig. 3** (a) The considered Li/Na ion migration pathways and (b) corresponding diffusion-barrier profiles for Li and Na on $\beta_{12}$ borophene.

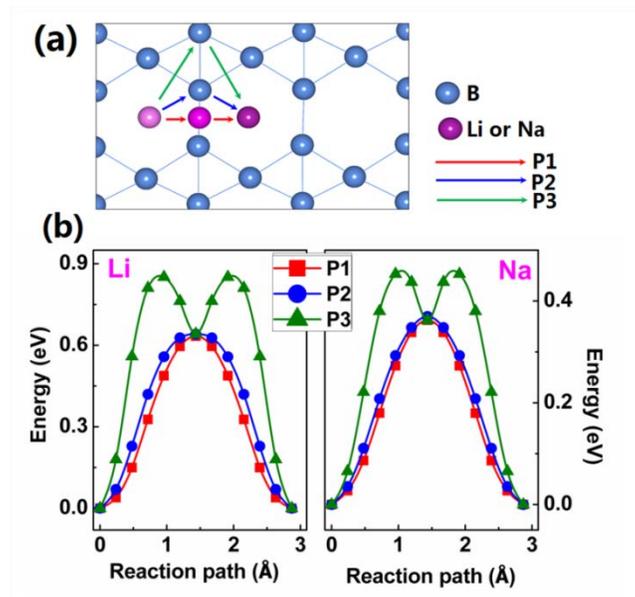

**Fig. 4** (a) The considered Li/Na ion migration pathways and (b) corresponding diffusion-barrier profiles for Li and Na on $\chi_3$ borophene.

Here we make further discussions on the diffusion characters of Li and Na ions on borophenes. For Li ion, $\beta_{12}$ and $\chi_3$ borophene both show the lowest diffusion barriers of ~0.6 eV. We note that the Li diffusion barrier on borophene is higher than typical 2D materials including graphene (~0.37 eV[17]), MoS$_2$ (~0.21 eV[19]) and most MXenes (0.05-0.15 eV[22-24]), suggesting a much lower LIB charge/discharge speed. However, it keeps comparable with some well-known anode materials such as TiO$_2$-based polymorphs (0.3–0.65 eV[49,50]), silicon (0.57 eV[51,52]), and phosphorene categories (0.13–0.76 eV[53,54]). On considering this, borophene still show great potential to serve as LIB electrodes. On the other hand, the diffusion barrier for Na ion is much lower, with the value of ~0.3 eV for both $\beta_{12}$ and $\chi_3$ borophene, which is well comparable to MoS$_2$ (0.28 eV[55]), silicence (0.25 eV[56]), and graphene-related electrodes (0.13-0.22 eV[57,58]). Therefore, high charge−discharge performances can be expected for $\beta_{12}$ and $\chi_3$ borophene to be used as electrodes for NIBs.

**Average open-circuit voltages and Li/Na storage capacities of borophene**

As we mentioned before, the storage capacity is a key indicator to judge the performance of electrode materials and is the current focus for improvement. To estimate the maximum storage capacity of Li and Na on $\beta_{12}$ and $\chi_3$ borophenes, we calculate the layer-resolved average adsorption energies for Li and Na adsorption layer by layer. Here we employ the $2 \times 2 \times 1$ supercell with increasing number of adsorbed Li and Na atoms on both sides of the host. The charge/discharge processes can be respectively described by the following half-cell reactions

$$B_{20} + xM^+ + xe^- \longleftrightarrow M_xB_{20} \quad (3)$$

for $\beta_{12}$ borophene, and

$$B_{16} + xM^+ + xe^- \longleftrightarrow M_xB_{16} \quad (4)$$

for $\chi_3$ borophene, where $M$= (Li, Na) and $x$ is the number of adsorbed atoms.

The layer-resolved average adsorption energies can be obtained by calculating the total energies difference before and after intercalation of a new Li/Na layer. To obtain more accurate results, both the atomic positions and lattice constants are fully relaxed for the configurations after intercalation. For Li/Na intercalation on $\beta_{12}$ borophene, the first layer of metal atoms is adsorbed above the vacancy ($S_\beta 1$ sites), which is the most stable Li/Na adsorption site with the lowest energy. For the second adsorption layer, we examine all the possibilities by assigning Li/Na atoms at $S_\beta 2$, $S_\beta 4$, and $S_\beta 11$, since the adsorption energies at these sites for a single Li/Na are quite close in our former calculations. Then we can obtain the average adsorption energy of each layer ($E_{ave}$), by using the following expression

$$E_{ave} = \left(E_{B_{20}M_{8n}} - E_{B_{20}M_{8(n-1)}} - 8E_M\right)/8 \quad (5)$$

where again M= (Li, Na) and $E_M$ is the cohesive energy in the bulk metal (Li/Na); $E_{B_{20}M_{8n}}$ and $E_{B_{20}M_{8(n-1)}}$ are the total energies of $\beta_{12}$ borophene with $n$ and ($n-1$) Li/Na adsorption layers. The number "8" in the formula represents eight adsorbed Li/Na atoms in each layer (for a $2 \times 2 \times 1$ supercell and on both sides). The average adsorption energy in the first layer is found to be -1.216 eV for Li and -0.858 eV for Na, respectively. For the second adsorption layer, we find that Li/Na can adsorbs at the $S_\beta 11$ sites, with a negative adsorption energy of -0.121 eV for Li and -0.062 eV for Na. However, when additional Li/Na layer is considered, the layer-resolved adsorption energy will become positive for all possible sites, which means that further adsorption of the third layer is energetically unfavorable. Therefore, as shown in Fig. 5(a) and 5(b), at most two Li and Na layers can be adsorbed for the $2 \times 2 \times 1$ $\beta_{12}$ borophene supercell, with corresponding chemical stoichiometry as $Li_4B_5$ and $Na_4B_5$, respectively. Then we can estimate the maximum capacity ($C_M$) by the following equation,

$$C_M = xF / M_{Borophene} \quad (6)$$

where $x$ represents the number of electrons involving the electrochemical process, $F$ derives from the Faraday constant with the value of 26798 mA h mol$^{-1}$, and $M_{Borophene}$ is the mass of borophene in g mol$^{-1}$. The calculated capacity of $\beta_{12}$ borophene is 1984 mA h g$^{-1}$ for both LIBs and NIBs. In addition, our results show that, during the intercalation of Li and Na ions on $\beta_{12}$ borophene, the lattice constants in the *xy* plane only experience slight change (about 1.1% and 2.2% tensile strain for Li and Na, respectively), which is another great advantage of $\beta_{12}$ borophene to be used as electrodes for rechargeable LIBs and NIBs. Further, we also estimate the average open-circuit voltage ($V_{ave}$), which is defined as

$$V_{ave} = \left( E_{B_{20}} + xE_M - E_{B_{20}M_x} \right) / xye \quad (7)$$

where $M$= (Li, Na), $E_{B_{20}}$ and $E_{B_{20}M_x}$ are the total energies of the $\beta_{12}$ borophene without and with Li/Na intercalation; $E_M$ is the cohesive energy of metal Li or Na; $y$ is the electronic charge of Li/Na ions in the electrolyte (here $y = 1$). For Li, the calculated average open-circuit voltage decreases from 1.26 to 0.61 V with increasing Li concentration from 8 to 16 atoms on $\beta_{12}$ borophene; while for Na, upon increasing the adsorption layers, the average open-circuit voltage will reduce from 0.88 to 0.42 V. On considering their low average open-circuit voltage values, $\beta_{12}$ borophene can be suitable to serve as anodes for both LIBs and NIBs.

Following the similar method, we estimate the maximum theoretical capacity of $\chi_3$ borophene. Again we employ a $2 \times 2 \times 1$ supercell in the calculation. For both Li and Na, we find that only one layer of atoms can be adsorbed, hence the maximum adsorption number is 8 for the $2 \times 2 \times 1$ supercell of $\chi_3$ borophene, which corresponds to the chemical stoichiometry of Li$_1$B$_2$ and Na$_1$B$_2$, respectively. Therefore, the calculated capacity of $\chi_3$ borophene for both LIBs and NIBs are 1240 mA h g$^{-1}$. In term of volume expansion, the lattice parameter of $\chi_3$ borophene only changes by about 0.9% during Li intercalation and 1.0% during Na intercalation. And the calculated average open-circuit voltage is found to be 1.09 V for LIBs and 0.78 V for NIBs. Therefore, $\chi_3$ borophene is also a good anode candidate material for LIBs and NIBs.

Finally we comment on the extremely high capacity of borophene for Li- and Na-storage, which can reach 1984 mA h g$^{-1}$ for $\beta_{12}$ borophene and 1240 mA h g$^{-1}$ for $\chi_3$ borophene. Fig. 5(c) shows the comparisons of the Li/Na capacities between borophene and other typical 2D anode materials, where some data are taken from literatures[16-29, 53-67]. To be noted, the theoretical capacity based on different calculating strategies may differ to some extent, as displayed in the supplementary information (see Tables S1 and S2). The comparisons in Fig. 5(c) follow the most common calculating method. As shown in Fig. 5(c), serving as LIB electrodes, the capacity of $\beta_{12}$ borophene can be 6-8 times of $MoS_2$/graphene (~335 mA h g$^{-1}$ [19]) and GeS nanosheet (256 mA h g$^{-1}$ [59]); 4-5 times of most MXenes, including $Mo_2C$ (526 mA h g$^{-1}$ [22]), $Nb_2C$ (542 mA h g$^{-1}$ [60]) and $Ti_3C_2$ (319 mA h g$^{-1}$ [61,23]); and more than 2 times of the recently reported Silicene (954 mA h g$^{-1}$ [27]) and $V_2C$ (941 mA h g$^{-1}$ [24]). Beside these 2D electrodes, the capacity of commercial graphite (~372 mA h g$^{-1}$ [3]) is also shown in Fig. 5 (c), which is found to be 5 times lower than $\beta_{12}$ borophene. For NIBs, the capacity of $\beta_{12}$ borophene can be 15 times of $MoS_2$ (146 mA h g$^{-1}$ [55]); more than 50 times of graphite (35 mA h g$^{-1}$ [62,63]); 4-6 times of most transition-metal dichalcogenides, such as $MoSe_2$ (513 mA h g$^{-1}$ [64]) and $TiS_2$ (339 mA h g$^{-1}$ [65]); 5-10 times of most MXenes, such as $V_2C$ (380 mA h g$^{-1}$ [24,61]) and $Nb_2C$ (271 mA h g$^{-1}$ [60]) and $Mo_2C$ (132 mA h g$^{-1}$ [22]); and also 2-4 times of the recently reported $Ca_2N$ (1138 mA h g$^{-1}$ [25]), phosphorene (865 mA h g$^{-1}$ [66]) and GeS nanosheet (512 mA h g$^{-1}$ [59]). For the other borophene, namely $\chi_3$ borophene, with a maximum capacity of 1240 mAh·g$^{-1}$ for both LIBs and NIBs, also shows considerable superiority compared with other 2D electrodes in terms of capacity. The extremely high capacity of borophene can be attributed to two major factors: (1) the small atomic mass of boron element; (2) the appearance of periodic voids in the 2D lattices which lowers the mass density and increase the (stable) adsorption sites. Moreover, for $\beta_{12}$ borophene, the mechanism of multilayer adsorption further boosts its storage capacity.

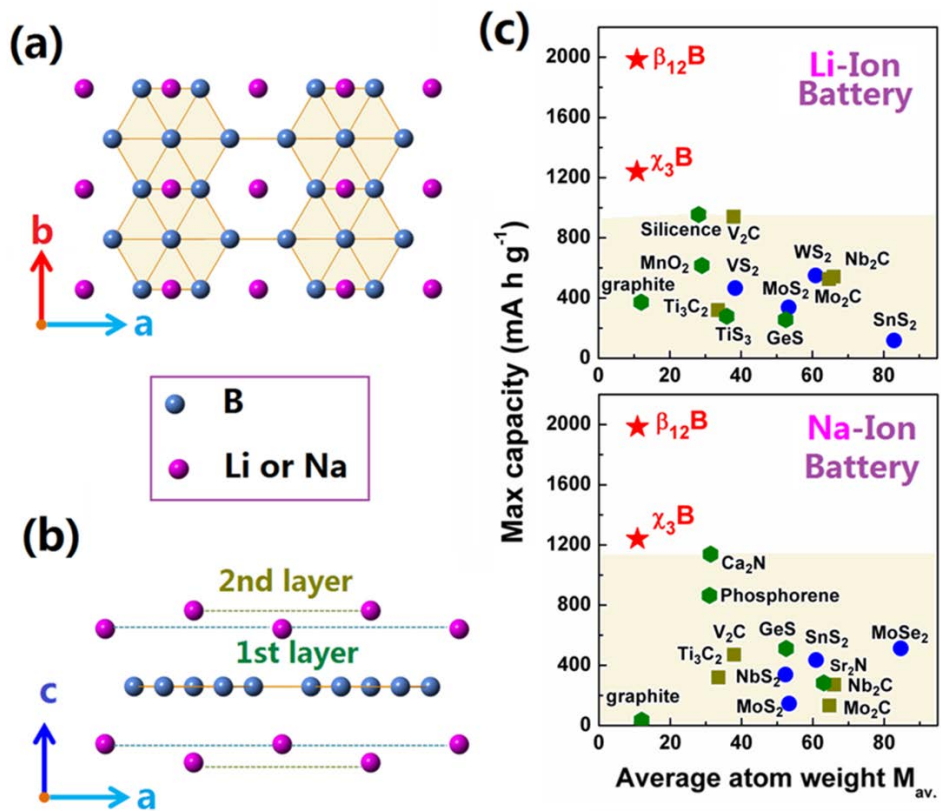

**Fig. 5** Top view (a) and (b) Side view of the atomic structure of Li-intercalated and Na-intercalated $\beta_{12}$ borophene, where at most two Li/Na adsorption layers can form. (c) Comparisons of the capacities between borophene and typical 2-D electrode materials for LIBs and NIBs. To be noted, the capacities in (c) are given for the electrodes materials in monolayer, except the case of few-layer graphene. And the 3-D commercial graphite is also listed for comparison. Some data are taken from literatures.[16-29, 53-67]

## 4 Summary

In summary, we have investigated the properties of borophene as potential electrode materials for LIBs and NIBs on the basis of first-principles calculations. Two experimentally stable borophene structures, namely $\beta_{12}$ and $\chi_3$ borophene, are studied. Our results show that borophene meets all the necessary requirements of a good electrode material: (1) metallic conductivities are maintained before and after adsorption; (2) low diffusion barriers, especially for Na ions, indicating good charge-discharge rates; (3) very small lattice change (< 2.2%) upon Li and Na intercalations, ensuring good cycling stabilities; (4) most importantly, both $\beta_{12}$ and $\chi_3$ borophene exhibit extremely high capacity for LIBs and NIBs, with 1984 mA h g$^{-1}$ for $\beta_{12}$ borophene and 1240 mA h g$^{-1}$ for $\chi_3$ borophene, which are several times higher than other typical electrode materials and is the highest among all 2D material based electrodes studied to date. Thus, we suggest that borophene is a highly promising electrode material for developing high performance LIBs and NIBs.

## Acknowledgments

This work is supported by Singapore Ministry of Education Academic Research Fund Tier 2 (MOE2015-T2-1-150), SUTD-SRG-EPD-2013062, and the National Natural Science Foundation of China (11574029, 11225418, 11504169 and 61575094).